\documentclass[11pt]{article}
    \usepackage{fullpage}

\title{How fast are algorithms reducing the demands on memory?  A survey of progress in space complexity}

\author{
 Hayden Rome\\
 \textit{MIT - CSAIL} \\
	Cambridge, MA, USA \\
 \texttt{hrome@alum.mit.edu}
  \and
 Jayson Lynch\\
 \textit{MIT - CSAIL} \\
 Cambridge, MA, USA \\
 \texttt{jaysonl@mit.edu}
 \and
  Jeffery Li\\
 \textit{MIT - CSAIL} \\
 	Cambridge, MA, USA \\
 	\texttt{jeli@mit.edu}
  \and
   Chirag Falor\\
  \textit{MIT - CSAIL} \\
	Cambridge, MA, USA \\
	\texttt{cfalor@mit.edu}
 \and
 Neil Thompson\\
 \textit{MIT - CSAIL} and \textit{IDE} \\
 	Cambridge, MA, USA \\
 	\texttt{neil\_t@mit.edu}
}

\usepackage{cite}
\usepackage{amsmath,amssymb,amsfonts}
\usepackage{algorithmic}
\usepackage{graphicx}
\usepackage{caption}
\usepackage{subcaption}
\usepackage{textcomp}
\usepackage{xcolor}
\usepackage{hyperref}
\usepackage{todonotes}
\def\BibTeX{{\rm B\kern-.05em{\sc i\kern-.025em b}\kern-.08em
    T\kern-.1667em\lower.7ex\hbox{E}\kern-.125emX}}

\begin{document}

\maketitle
\begin{abstract}

Algorithm research focuses primarily on how many operations processors need to do (time complexity). But for many problems, both the runtime and energy used are dominated by memory accesses. In this paper, we present the first broad survey of how algorithmic progress has improved memory usage (space complexity). We analyze 118 of the most important algorithm problems in computer science, reviewing the 800+ algorithms used to solve them. 
Our results show that space complexity has become much more important in recent years as worries have arisen about memory access bottle-necking performance (the ``memory wall''). In 20\% of cases we find that space complexity improvements for large problems (n=1 billion) outpaced improvements in DRAM access speed, suggesting that for these problems algorithmic progress played a larger role than hardware progress in minimizing memory access delays. Increasingly, we also see the emergence of algorithmic Pareto frontiers, where getting better asymptotic time complexity for a problem requires getting worse asymptotic space complexity, and vice-versa. This tension implies that programmers will increasingly need to consider multiple algorithmic options to understand which is best for their particular problem. 
To help theorists and practitioners alike consider these trade-offs, we have created a reference for them at \href{https://algorithm-wiki.csail.mit.edu}{Algorithm Wiki}.

\end{abstract}

\section{Introduction}\label{sec:introduction}

In this work, we aim to provide the first comprehensive survey of the space complexities of algorithms across a wide range of important problems in computer science. Previous work by Sherry and Thompson~\cite{yash} gave a similar survey on time complexities. In their work, they identified important problems in computer science and analyzed how fast algorithms improve, comparing that rate with the rate of improvement of hardware processing speed. The scope of this paper is the analysis of space complexities of algorithms from Sherry and Thompson's work~\cite{yash} as a whole, rather than discussing the specific individual space complexities of algorithms. 
We also provide our results in an easy-to-access repository, \href{https://algorithm-wiki.csail.mit.edu}{Algorithm Wiki}, for the ease of theorists and practitioners alike.

The broad scope and comprehensiveness of our study allows us to analyze (i) the overall distribution of best space complexities across algorithm problems, (ii) the growing importance of space complexity in the algorithm literature, (iii) how space complexity and time complexity correlate across problems, (iv) the overall improvement rates of space complexity over time, and (v) the emergence of trade-offs between time and space complexity (i.e. Pareto frontiers).

The distributions of the best space and time complexities of algorithms are summarized in Figure~\ref{fig:algorithm_landscape} in Section~\ref{subsec:algorithm_landscape}. One of the most notable features is that 84\% of algorithm problems have linear or smaller space complexity, implying limited area for improvement. The algorithm problems with larger-than-linear space complexity represent a much smaller fraction of all algorithms (16\%) than is true for time complexity -- in agreement with common belief that most algorithms will have space complexity be no larger than time complexity. 

In the articles that describe new algorithms, we find that only 14\% of papers gave an analysis of the space complexity of their algorithms. A further 9\% are analyzed in later works. The share of papers doing space complexity analysis has been increasing over time, suggesting that algorithms researchers are considering this an increasingly important topic.

We investigate the correlation between individual algorithms time and space complexities in Section~\ref{subsec:problems_time_v_space}. This may help us understand what types of algorithms admit space-efficient versions, and also direct us to cases where improvements in the space-complexity seem likely to be obtainable. This will hopefully help researchers find and prioritize open problems.

We calculate the yearly percentage improvement rate from the space complexity improvements in Section~\ref{subsec:rates}. We find only about 20\% of algorithms had meaningful improvements. However, of those improvements that did occur, the rate was often larger than the rate of hardware improvements to DRAM. We also see evidence that space complexity improvements did not keep pace with time complexity improvements - potentially exacerbating ``memory wall" challenges where memory accesses bottleneck performance.

Now that we are investigating two dimensions of algorithm performance, both time and space complexity, there can be non-trivial trade-offs between different algorithms for a given problem. We find that 17\% of our problems have such a time-space tradeoff, discussed in Section~\ref{subsec:tradeoffs}. Interestingly, when we view this historically, the fraction of families with such a tradeoff has been increasing over time.

This paper is organized as follows. In Section~\ref{sec:relatedwork}, we discuss previous algorithmic surveys. In Section~\ref{sec:methods}, we describe the methodology we used to survey and categorize results as well as derive missing space complexities. In Section~\ref{sec:results}, we present the main results of the analysis of space complexities. In Sections~\ref{sec:conclusion}, we discuss some potential implications of this work.

\section{Related Work}\label{sec:relatedwork}

The concept of space complexity was introduced in 1965 by Stearns \cite{Stearns1965}. Since then, space complexity has been an important area of research in algorithm analysis, alongside time complexity. While there have been some surveys of space complexity, such as Michel's work \cite{Michel1992_space_comp_survey}, these have tended to focus on general theorems and relationships between models of computation, rather than the space complexity of specific algorithms. Some textbooks, such as Alfred~\cite{Alfred1974} and CLRS~\cite{CLRS}, have provided analyses of space complexity for fundamental algorithms, but they are limited in scope and do not cover the full algorithmic history of the problems, which we aim to do in this work.

Leiserson et al.~\cite{roomatthetop} argue that progress in computing will come from three sources: hardware, software, and algorithms. In this work, we focus on the contributions from algorithms. The algorithmic progress on time complexity was already studied by Sherry and Thompson~\cite{yash}, which defined 113 key algorithm problems prioritized by the field, and examined progress. In that paper, they look at improvements in time complexity for 113 different \emph{problem families}. In this paper, we turn the focus from time complexity to space complexity. How space-efficient are the different algorithms for a problem family \footnote{Each algorithm problem has an associated family of algorithms that solve it. An example of a problem family is `Comparison sorting', and an algorithm that would belong to that problem family is `Merge Sort'.}, and how quickly do we see new algorithms that are more space-efficient than previous algorithms for a specific algorithm family?

Work by Liu et al.~\cite{optimal, liu2021metastudy} took the same database of algorithms and problem families as in Sherry and Thompson~\cite{yash} and surveyed results on lower bounds for the time complexity of these problem families. They found 64\% of the surveyed algorithms are optimal, meaning their asymptotic lower bounds matched their upper bounds and so no more asymptotic improvements are possible. This gives a sense of how much of the possible algorithmic progress has already been achieved, and where we will need to turn to other techniques like approximation algorithms and parallelization to see further algorithmic gains.

In this paper, we use the same sampling of algorithms and problem families as in these two algorithmic survey works. These papers and other large algorithmic surveys do not address the space complexities of these algorithms and how they progress. As part of this paper's contribution, we have derived and gathered the space complexities of more than 800 algorithms from this sample--many of which the original papers did not even mention their algorithm's space complexity.

\section{Methods}\label{sec:methods}

In what follows, we take inspiration from Sherry and Thompson's work ~\cite{yash} on methodology because it allows a direct comparison between time and space complexity. Many of the following formulas and categorizations derive from their methods.

\subsection{Gathering Algorithms}\label{subsec:gatheringalgorithms}

Sherry and Thompson~\cite{yash} gathered algorithms from 57 textbooks and more than 1,137 research papers, arguing that their broad coverage over time and sub-fields of computer science made their study quite comprehensive of important algorithms. For our analysis here, we omit approximation algorithms, quantum algorithms, parallel algorithms, distributed algorithms, and algorithms for inexact problems (see Section~\ref{sec:background}). We omit these algorithms because they are analyzed with different metrics and they don't provide exact like-to-like comparisons with exact algorithms for exact problems. In addition to the aforementioned algorithm types, we omit a few algorithms for which we were unable to find or access a source.

\subsection{Grouping Algorithms}\label{subsec:groupingalgorithms}

We group algorithms into \emph{problem families}, meaning algorithms that solve the same problem. As in that work, we report on the canonical problem within each family, but in our data repository we also report how on problem variations, for example all-pairs shortest path (APSP) variations for directed or undirected graphs, weighted or unweighted graphs, and dense or sparse graphs. Clarifying some classification ambiguities in prior work has lead us to add an additional five algorithm families; so our analysis compares a total of 118 different algorithm families. 

\subsection{Model of Computation}\label{subsubsec:modelofcomputation}

The majority of algorithms analyzed use either the \emph{word RAM} model of computation or the \emph{real RAM} model. The word RAM model is a random-access machine (RAM) that operates on words, or groups, of bits \cite{wordram}. These words are typically of size $O(\log n)$. The space complexity in this model is typically measured in number of words rather than bits. The real RAM model operates on real numbers exactly and each memory cell contains a real number of arbitrary precision and size\cite{realram}. 

Another important consideration regards how to count memory in the RAM model. Two natural methods are 1) to count the number of unique memory cells that have been written to, and 2) to look at how many memory cells are between the smallest and largest numbered memory cells that have been written to. We use the second method, the high-water mark measure. It is worth noting that in the Word RAM model, when we say \emph{constant}--or $O(1)$--auxiliary space, this means a constant number of $\log n$-sized words.

A small number of algorithms' analysis uses other models of computation, such as Turing machines. We note the computation model for each algorithm in our data tables.

\subsection{Auxiliary Space Complexity}\label{subsec:auxiliary}

There are multiple different space complexities, or memory requirements, that an algorithm has. These include: input space complexity, output space complexity, auxiliary space complexity, and total space complexity. Input space complexity is the size of the input (typically, the other space complexity measures are with respect to this). Output space complexity is the size of the output. Auxiliary space complexity is the amount of space required during the computation that excludes the input and output. For example, if an algorithm requires constructing a matrix that is not part of the output, then the matrix would count towards the auxiliary space. The total space complexity is the sum of the input, output, and auxiliary space complexities. 

When discussing the space complexity of algorithms, we consider only the \emph{auxiliary space} because it provides a fine-grained view of the space that is actually needed. Furthermore, since all algorithms for the same problem will have the same input and output space complexities, focusing on auxiliary space complexity allows us to more easily distinguish space-efficient algorithms from space-inefficient algorithms.

\subsection{Asymptotic Complexities}\label{subsec:asymptotic}
\subsubsection{Quantizing}\label{subsec:quantizing}

To analyze the improvements going from one time complexity to another, we classify algorithms' time and space complexity using eight quantized classifications. In order, these are: 
\begin{itemize}
	\item Constant -- $O(1)$
	\item $\log n$ -- $O(\log^c n)$ for some constant $c > 0$
	\item Linear -- $O(n)$
	\item $n \log n$ -- $O(n \log^c n)$ for some constant $c > 0$
	\item Quadratic -- $O(n^2)$
	\item Cubic -- $O(n^3)$
	\item Supercubic -- $O(n^c)$ for some constant $c > 3$; also, super-polynomial but sub-exponential
	\item Exponential/factorial -- $O(c^n)$ for some constant $c > 1$ or $O(n!)$
\end{itemize}

For example, if we have an algorithm whose space complexity is $O(n^{2.5})$, we would classify this as between quadratic and cubic, and thus round it up to cubic. In other cases we use intermediate classifications.

\subsubsection{Multiple Parameters}\label{subsec:parameters}

Some problems have only one main parameter $n$ which is the focus of the time and space complexities. However, there are other problems where there are multiple parameters that show up in these complexities. For our analysis, we need a single overall classification for the algorithms that solve these problems. To do this we determine a single asymptotic formula to use as the basis of the classification, which is typically the input size.

For the majority of the problems, the formula is a single variable (e.g. $n$). For matrix problems with $n$ and $m$ being the number of rows and columns respectively, this formula is $n^2$ for square matrices or $mn$ for rectangular matrices. For graph problems, we often standardize to the number of vertices $V$, and when we consider dense graphs we plug in $O(V^2)$ for the number of edges $E$. Note that for our analysis we normalize to have inputs of size "$n$", so we would say standard Matrix Multiplication runs in time $n^{1.5}$ in the number of matrix entries although it is more common in computer science to say the running time is cubic in the number of rows and columns. Similarly for the standard algorithm for All-Pairs Shortest Path in dense graphs is considered to run in time $n^{1.5}$ where $n= |V|+|E| = |V|^2$ although it is more common to say APSP runs in $|V|^3$ time.

\subsection{Rate of Improvement Calculation}\label{subsec:methods_rates}

We calculate the improvement rate of space complexity for problem families as follows. An improvement from algorithm $i$ to algorithm $j$ is calculated as
\begin{align}
	\text{Improvement}_{i\rightarrow j} = \frac{\text{Words}_i(n)}{\text{Words}_j(n)}
\end{align}
where $n$ is the problem size and the number of words (units of memory in the Word RAM model) is calculated using the asymptotic auxiliary space complexity. For this work, we consider problem sizes of $n=10^3, 10^6, 10^9$.

Thus, to estimate the average per-year percentage improvement, we calculate:

\begin{align}
	\text{YearlyImprovement}_{i\rightarrow j} =  \left( \frac{\text{Words}_i (n)}{\text{Words}_j(n)} \right)^{1/t} - 1
\end{align}
We only consider years since 1940 to be those where an algorithm was eligible for improvement.

When calculating the rates of improvement using multiple parameters, we use the same ratios of parameters as in Sherry and Thompson's work~\cite{yash} in order to best draw like-to-like comparisons.

\section{Results}\label{sec:results}

\subsection{A Look at the Algorithm Research Landscape}\label{subsec:algorithm_landscape}

\begin{figure*}
	\centering
	\includegraphics[width=.95\textwidth]{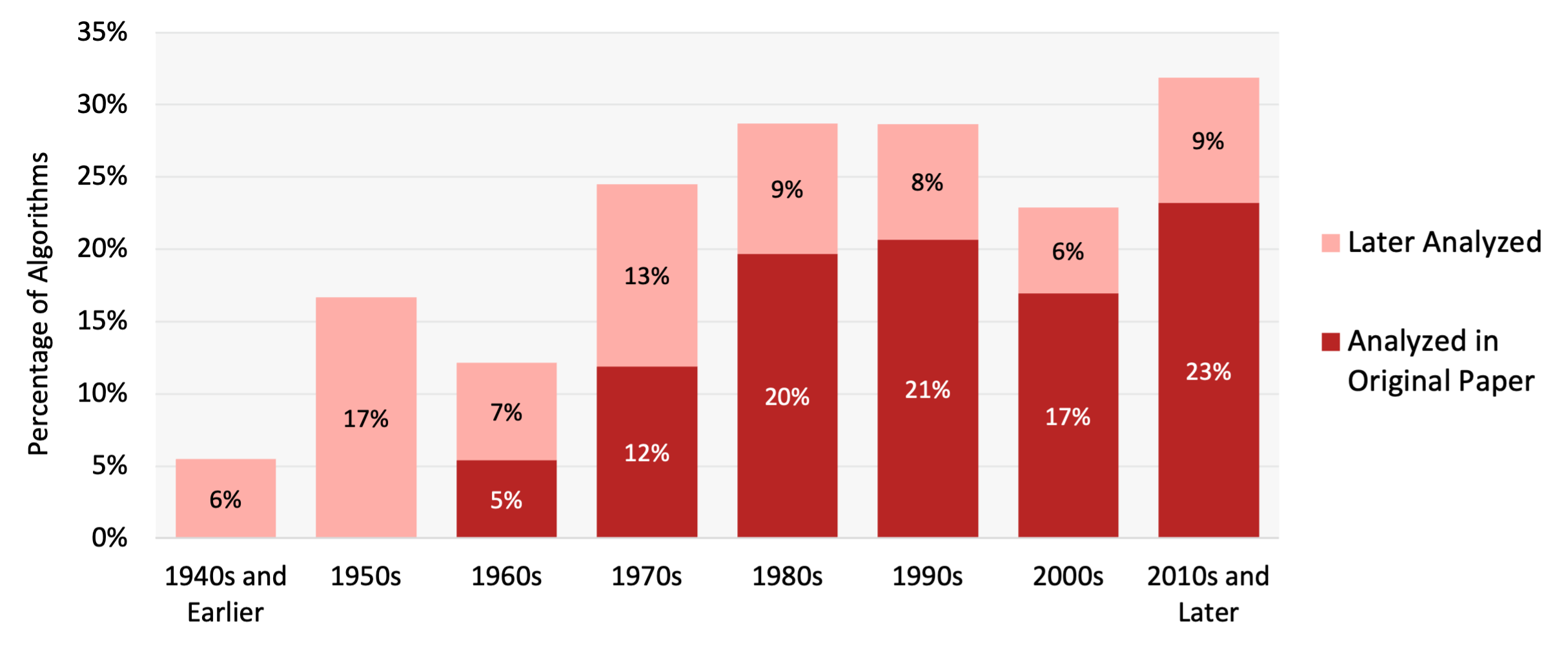}
	\caption[Papers with Space Analysis]{Percentage of algorithms published in each decade with space analysis in the original paper or later.}
	\label{fig:space_analysis}
\end{figure*}

Given the importance of space complexity, it is remarkable how rarely it is derived in algorithm articles, as Figure~\ref{fig:space_analysis} shows. Since the the notion of space complexity was only formally introduced in 1965, it is not surprising that earlier articles omitted the topic. But in the decades since, the share has grown steadily, but slowly, by 3.6\% per decade. A plausible explanation for this is that space complexity is becoming more important as the size of real-world inputs grows and more and more problems become memory bound. By the 2010s, nearly  a third of algorithm articles had space complexity derived (either by the original paper or subsequent work).

\begin{figure*}
	\centering
	\begin{subfigure}[b]{.95\textwidth}
		\includegraphics[width=\textwidth]{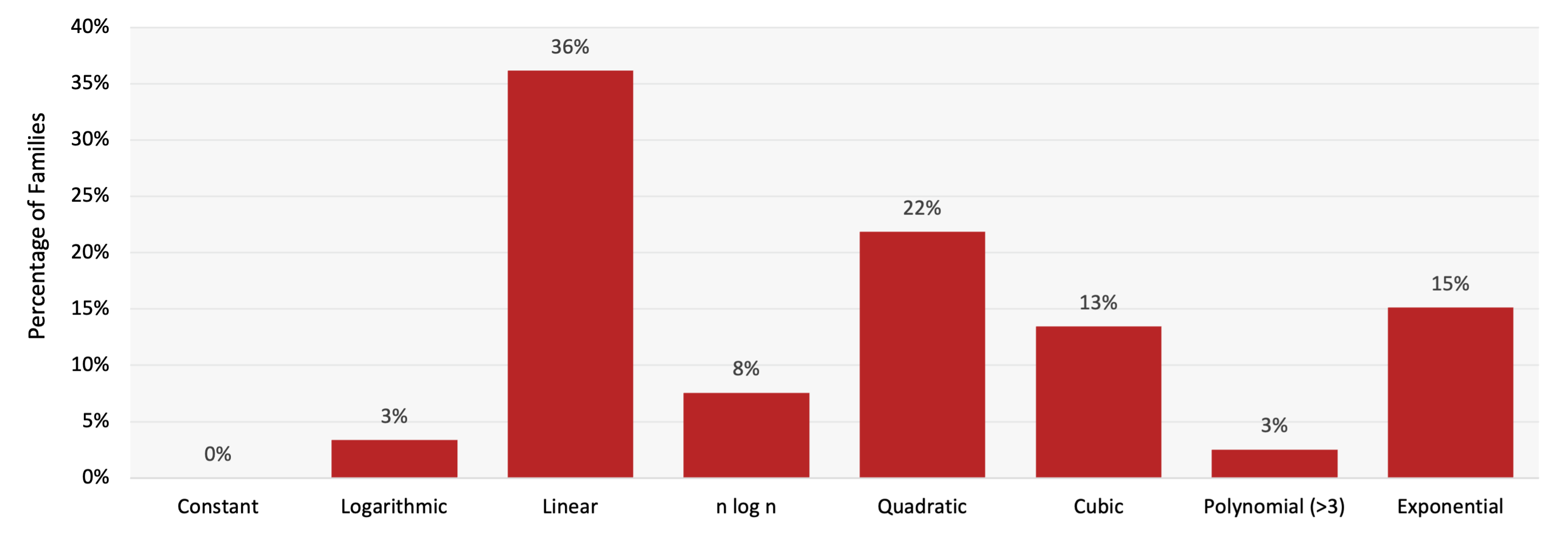}
		\caption{Problem families' current best time complexity}
		\label{subfig:time_landscape}
	\end{subfigure}
	\\
	\begin{subfigure}[b]{.95\textwidth}
		\includegraphics[width=\textwidth]{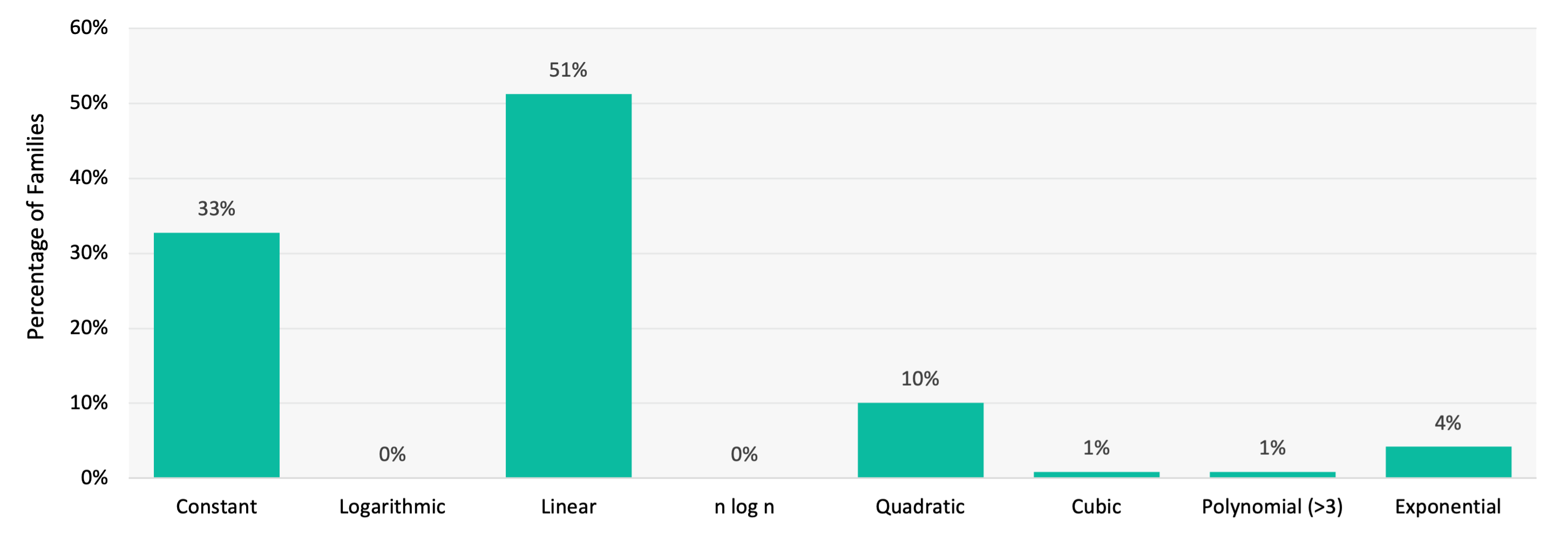}
		\caption{Problem families' current best auxiliary space complexity}
		\label{subfig:space_landscape}
	\end{subfigure}
	\caption{Overview of Algorithm Landscape}
	\label{fig:algorithm_landscape}
\end{figure*}

We show the distribution of the best time complexities of problem families in Figure~\ref{subfig:time_landscape} and the best space complexities in Figure~\ref{subfig:space_landscape}. Here we see over 80\% of algorithms have auxiliary space complexity that is linear or smaller; this is a much larger fraction than the 39\% of algorithms that have a linear or smaller time complexity.  Generally, we expect and see that the distribution of space complexities is concentrated towards lower asymptotic complexities more than are the time complexities; because the whenever a value is stored, it takes at least one time step to store it. 
For example, if we need to store a size $O(n)$ array, we need to perform at least $O(n)$ steps--at least one step for each element. All of these storage time steps count towards the time complexity of the algorithm.

Interestingly, though, we do not see any problems in which the best space complexity is poly-logarithmic or quasilinear $O(n \log^c n)$ for any constant $c > 0$, whereas we see more than 10\% of time complexities fall into these categories. This may be a result of the model in which we are considering being the word RAM model with words of size $O(\log n)$- bits because $\log n$ bits are needed to store an index into $n$ items.

We also see that 12\% of algorithm families have polynomial space complexities of at least $\Omega(n^2)$. If we set constants to be $1$ and word size to be 64 bits, then problems of size 64,000 would fill a typical processor's 32 GB RAM and a problem of size 360,000 would fill up a 1TB hard-drive.  Similarly, 4\% of problems have exponential space complexity, so that problems of size 37 would fill such memory. In these cases, we see the importance space complexity improvements for making large data problems computationally efficient.

\subsection{Comparing Individual Problems' Space and Time Complexities}\label{subsec:problems_time_v_space}

\begin{figure*}[h]
	\includegraphics[width=.9\textwidth]{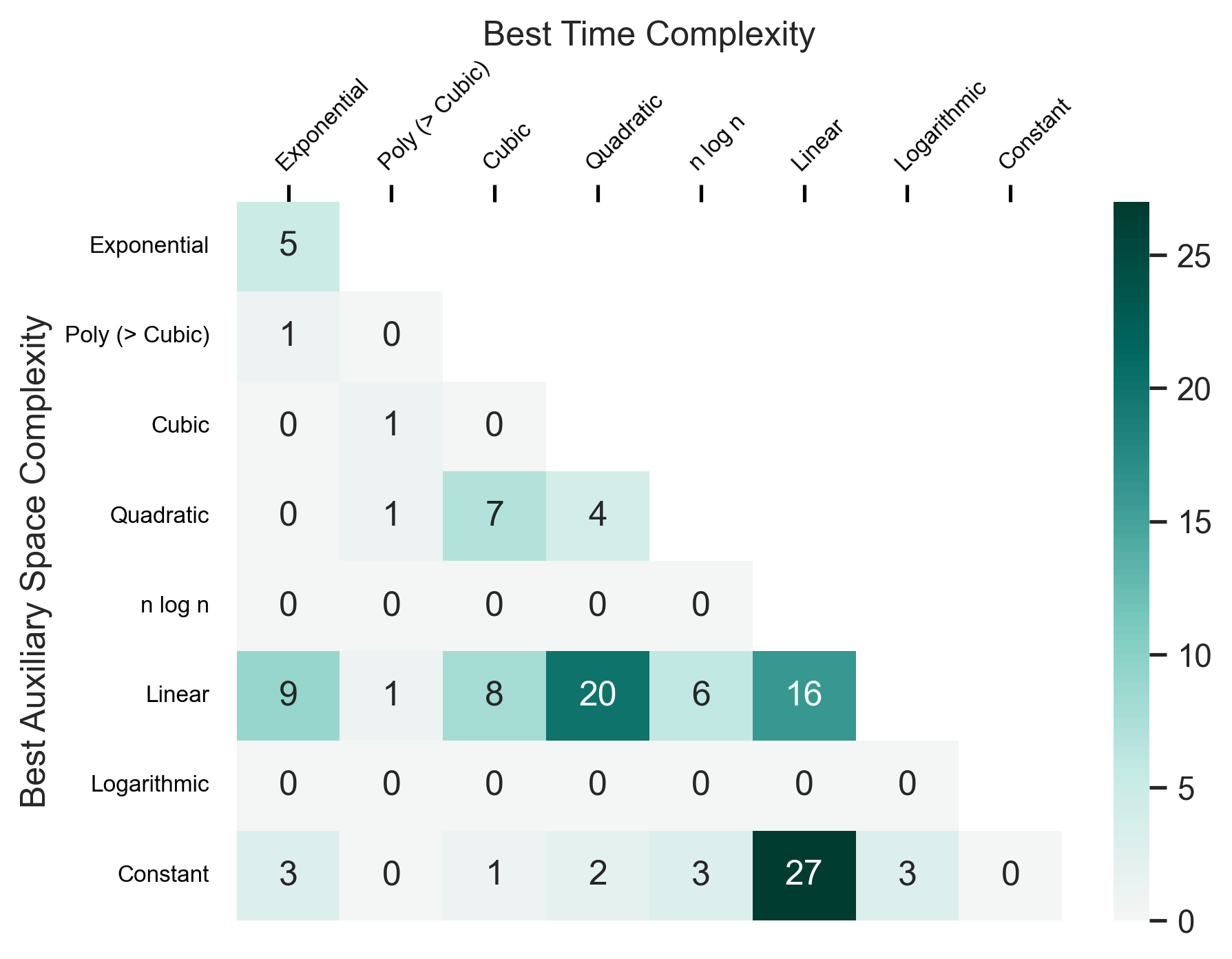}
	\caption[Comparing Problems' Best Space and Time Complexities]{Heat map of the distribution of the best space and best time complexities for 118 problem families.}
	\label{fig:heatmap_space_vs_time}
\end{figure*}

In addition to looking at the broad distributions of problems' best space and time complexities, like in Figure~\ref{fig:algorithm_landscape}, we can look at which specific space and time complexities occur together. Figure~\ref{fig:heatmap_space_vs_time} is a heat map of the best auxiliary space and time complexities for each problem family.

We can see that there tend to be more constant and linear space families as the best time complexity gets smaller.
It is interesting to note that the two largest clusters (27 and 20) are cases where the time complexity is a linear factor larger than the space complexity, implying that each space in memory is being used for $n$ computations, on average.

Our analysis also highlights 10 algorithm families that are likely to see algorithmic progress once researchers focus on them because of pervasive success with reducing polynomial space complexities.
These are the algorithms in the top left corner below the diagonal and above linear space complexity. 
For the algorithm community, we provide a list of these algorithms (as well as all other problems that have best auxiliary space greater than linear in Appendix~\ref{sec:superlinear_space}.

\subsection{The Rate of Space Complexity Improvement}\label{subsec:rates}

\begin{figure*}
    \centering
    \begin{subfigure}[b]{0.85\textwidth}
        \includegraphics[width=\textwidth]{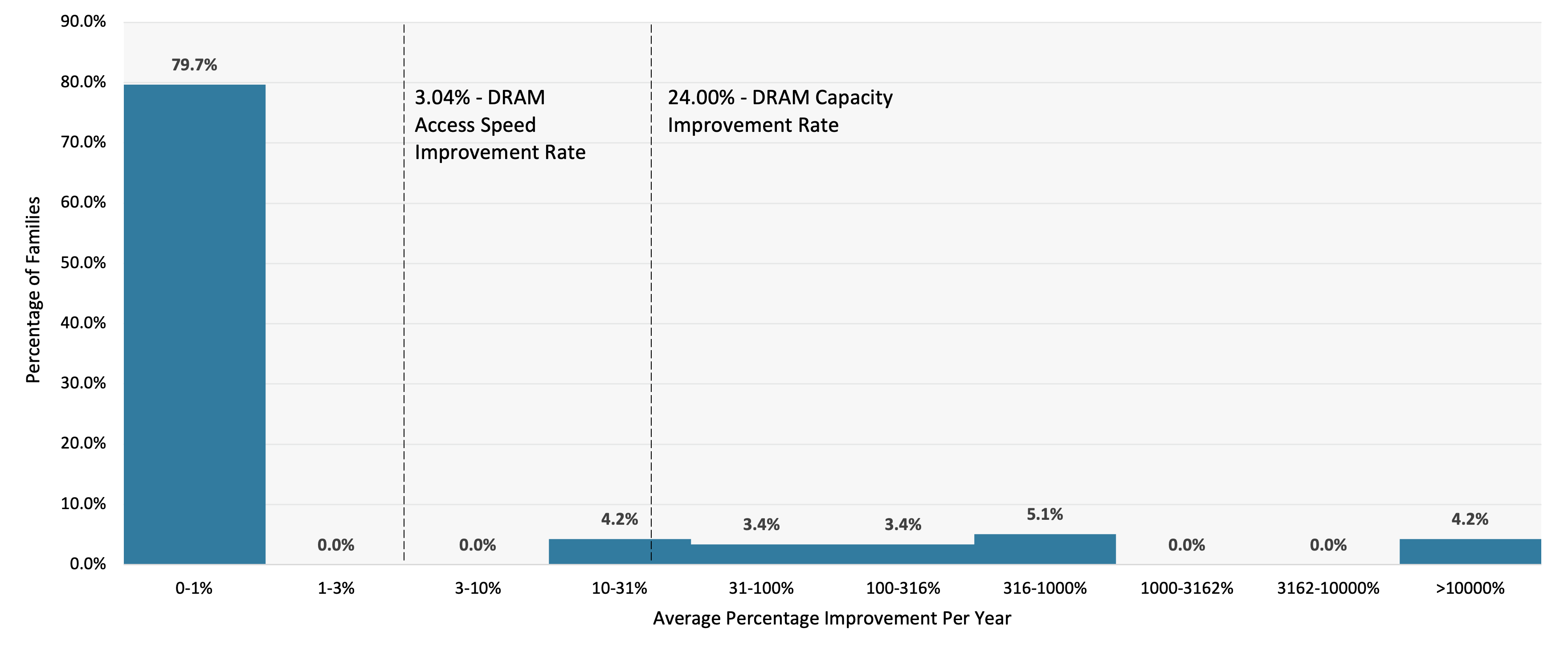}
        \caption{$n=1,000$}
        \label{subfig:rate10^3}
    \end{subfigure}
    \\
    \begin{subfigure}[b]{0.85\textwidth}
        \includegraphics[width=\textwidth]{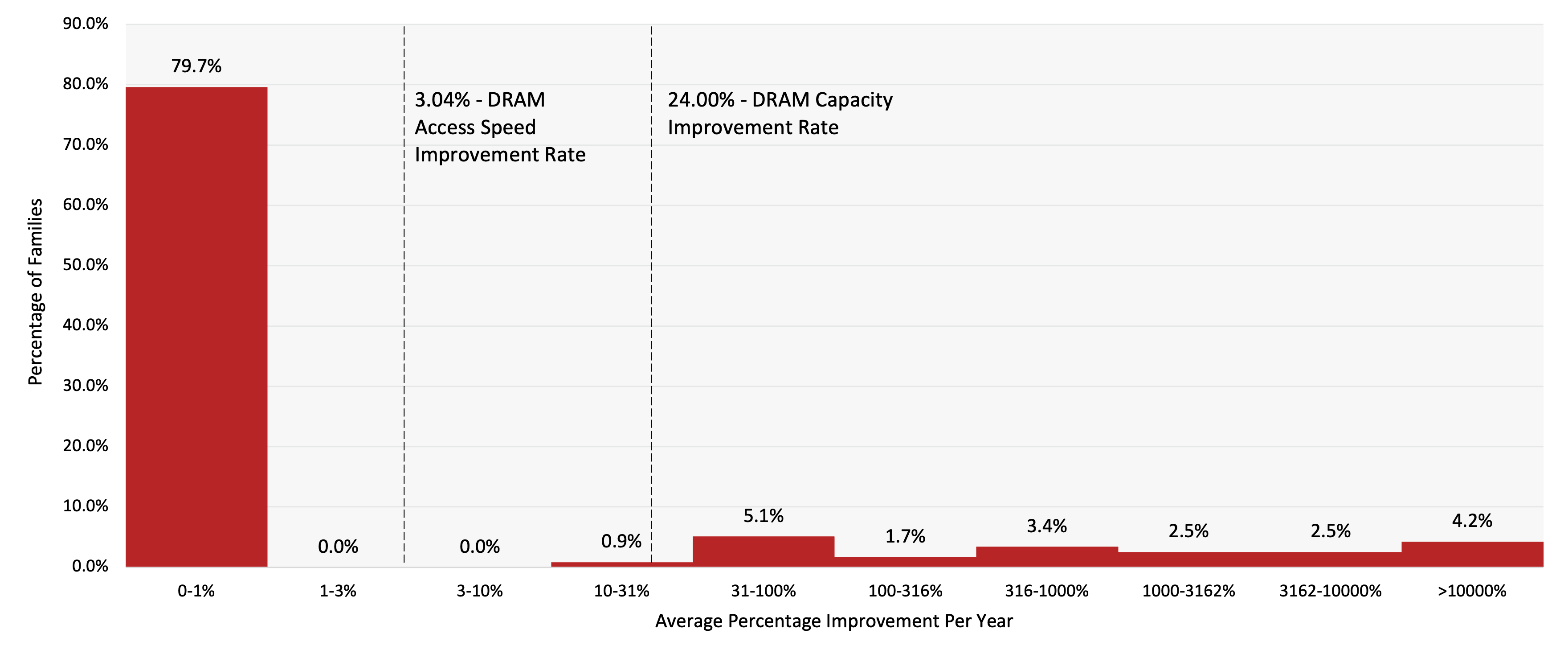}
        \caption{$n=10^6$}
        \label{subfig:rate10^6}
    \end{subfigure}
    \\
    \begin{subfigure}[b]{0.85\textwidth}
        \includegraphics[width=\textwidth]{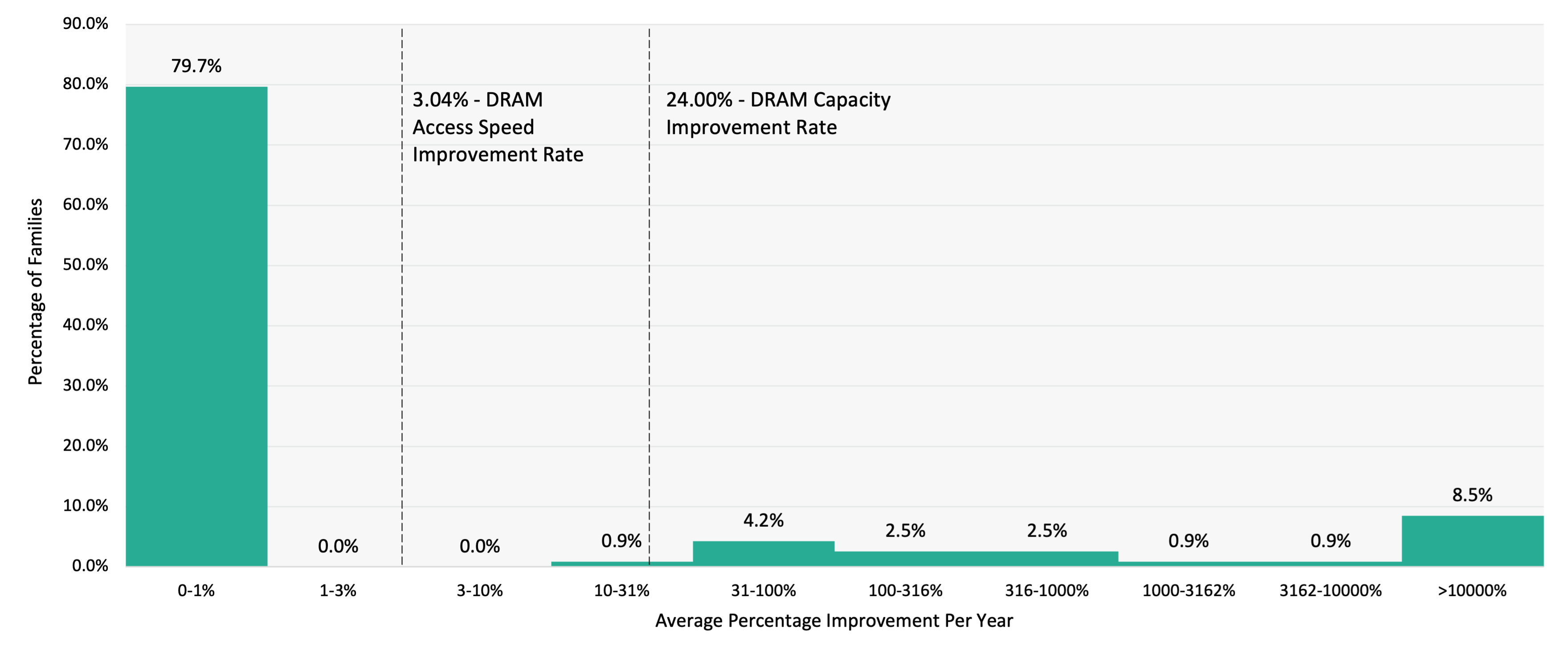}
        \caption{$n=10^9$}
        \label{subfig:rate10^9}
    \end{subfigure}
    \caption{Distribution of average yearly improvement rates for 118 problem families, as calculated based on asymptotic space complexity, for problems of size: (a) $n =$ 1 thousand, (b) $n = $ 1 million, and (c) $n =$ 1 billion. Both DRAM yearly improvement rates were calculated using DRAM data from Hennessy  and Patterson~\cite{computerarchitecture}.}
    \label{fig:rates}
\end{figure*}

Here, we analyze the improvement rate of the space complexity of problem families. For each problem family, we look at the very first algorithm published, and calculate the annualized rate of improvement from that algorithm until the current best performing one. These average annualized improvement rates are what is plotted in the histogram in Figure~\ref{fig:rates}, for multiple different input sizes.

Overall, the improvement rate of space complexity is generally less than the improvement rate of time complexity found in \cite{yash}. That paper calculated, for problem sizes of 1,000,000 the median rate of improvement for algorithm speed was 15\% and that 30\% of algorithms had rates of speed improvements greater than hardware improvements. By contrast, for space complexity, we see that nearly 80\% of algorithms families saw no asymptotic improvements in space since their original algorithm was published, suggesting it is either harder to make progress in space complexity or easier to find an optimal space complexity in an initial analysis.

While only ~20\% of algorithm families have space complexity improvements, the size of these gains is highly relevant to questions of memory access bottlenecks (Memory Wall) arising because processing speed is improving faster than the memory access speed\cite{wulf1995hitting, mckee2004reflections}. From data in Hennessy and Patterson~\cite{computerarchitecture} we estimate that DRAM access speed has improved by 3\% per year since 2000 and DRAM capacity has increased by 24\% per year. We find that, for all the problem sizes we analyze, space complexity improvements have outpaced DRAM access speed improvements and most of them have even improved faster than DRAM capacity. This is important because sometimes a smaller space complexity indicates fewer memory accesses needed, implying that algorithms are reducing the need for more memory faster than hardware is increasing its availability.\footnote{The reason that this is not always the case, is because one can make many data accesses to a single position in memory, so the number of accesses and the space used can differ.} A second benefit of these lower memory needs is that data can sometimes be moved up the memory hierarchy (e.g. to cache), significantly speeding up the accesses for that data. \footnote{Some more detailed computational models capture the notions of temporal and spatial locality in algorithms' memory access directly, which better predicts efficient use of computer cache\cite{aggarwal1988input}.}

\subsection{Time-Space Tradeoffs}\label{subsec:tradeoffs}

When choosing an algorithm to solve a given problem, the obvious choice is to choose the algorithm with the best asymptotic time and space complexity. However, some problems may not have one single ``best" algorithm for both. Instead, there may be multiple algorithms with a tradeoff in time and space complexity. Thus, there is a Pareto frontier of algorithms that are all reasonable, depending on how much one cares about space efficiency or speed.

\begin{figure*}
    \centering
        \begin{subfigure}[b]{0.95\textwidth}
        \includegraphics[width=\textwidth]{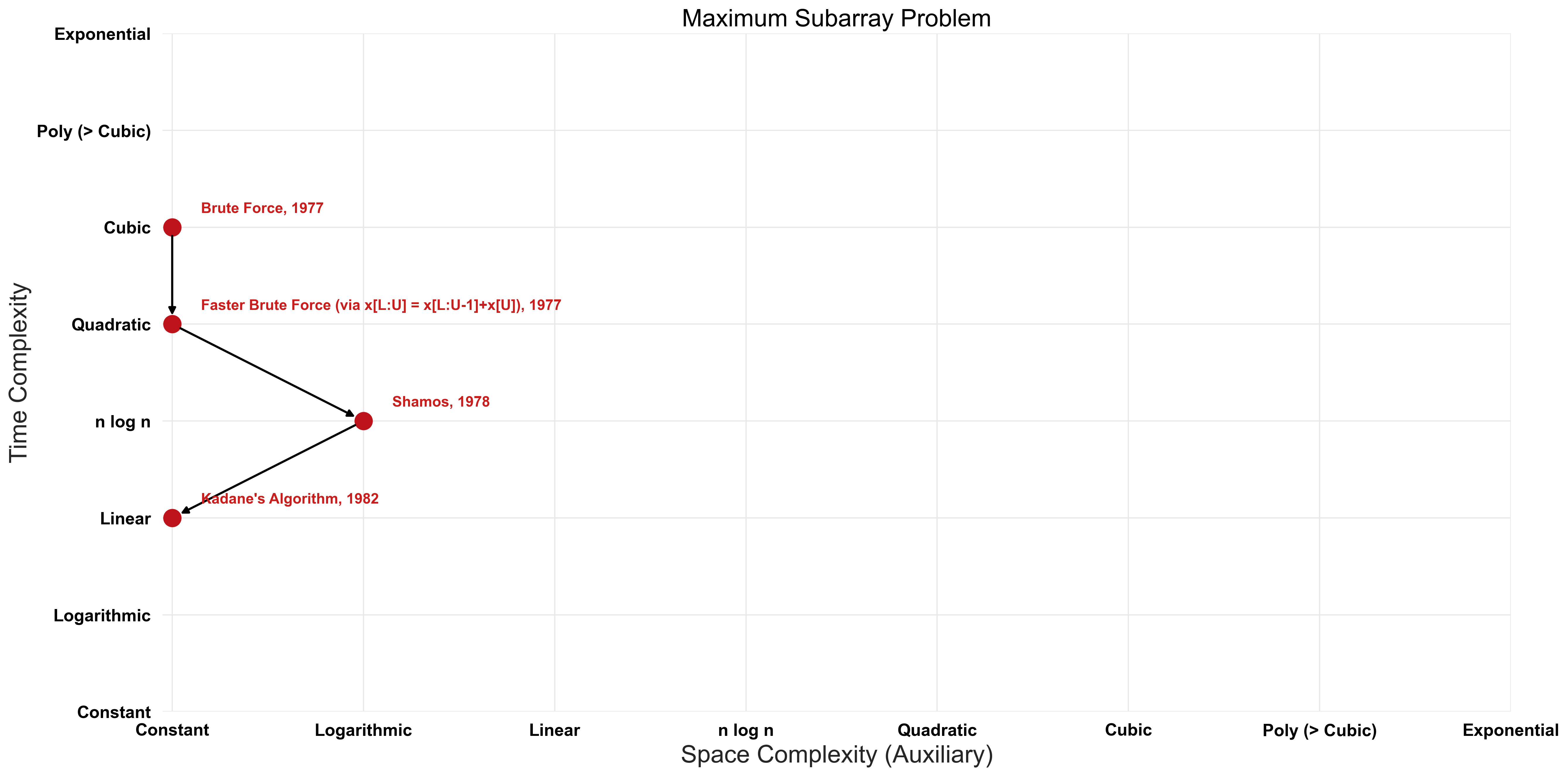}
        \caption{The evolution of the Pareto frontier for the Maximum Subarray Problem family.}
        \label{subfig:pareto_example}
    \end{subfigure}
    \\
    \begin{subfigure}[b]{0.95\textwidth}
        \includegraphics[width=\textwidth]{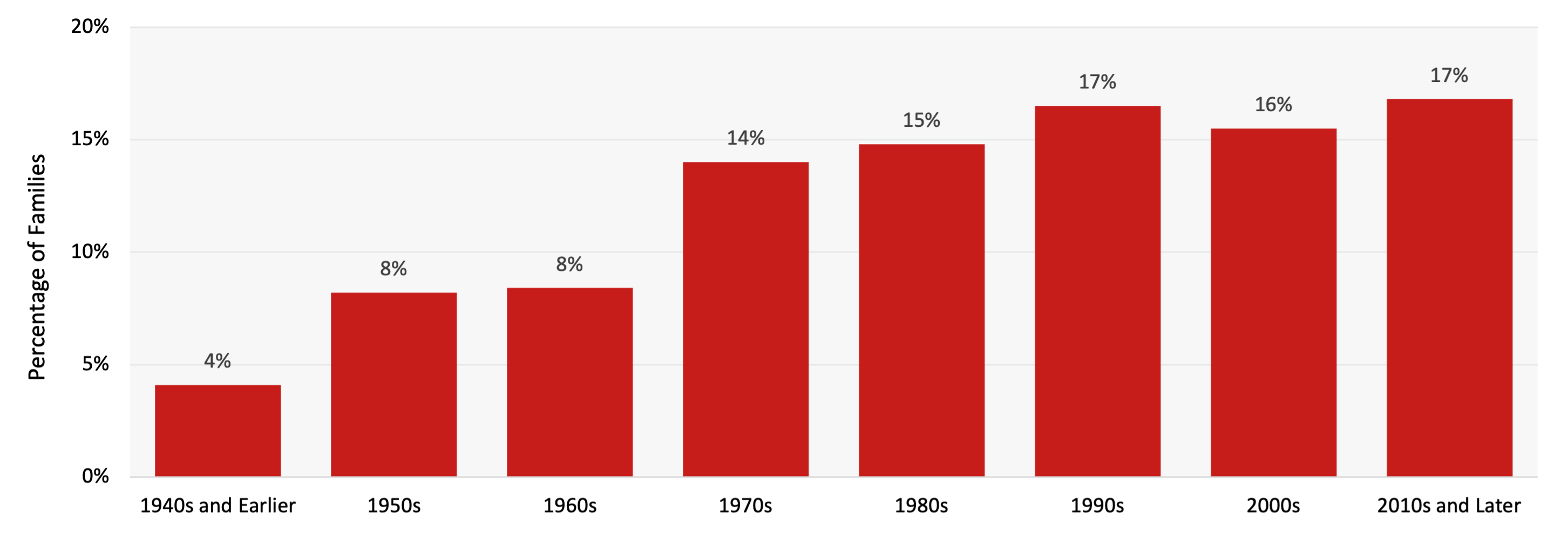}
        \caption{Percentage of problem families in each decade that have a time-space tradeoff between algorithms on the Pareto frontier.}
        \label{subfig:tradeoff_history}
    \end{subfigure}
    \caption{Time-Space Tradeoffs}
    \label{fig:tradeoffs}
\end{figure*}

We can visualize the progress of specific problem families over time, like for the Maximum Subarray Problem family in Figure~\ref{subfig:pareto_example}. The figures we produce like this include all of the algorithms that, at some point, were on the Pareto frontier for the problem family. This figure shows that there was no tradeoff necessary until 1978, when Shamos' algorithm was published with a slightly worse auxiliary space complexity but better time complexity than the best algorithm at the time. Then, the tradeoff disappeared in 1982 with Kadane's algorithm, which is still the single optimal algorithm for this problem. We can visualize the progress of specific problem families over time, like for the Maximum Subarray Problem family in Figure~\ref{subfig:pareto_example}. The figures we produce like this include all of the algorithms that, at some point, were on the Pareto frontier for the problem family. This figure shows that there was no tradeoff necessary until 1978, when Shamos' algorithm was published with a slightly worse auxiliary space complexity but better time complexity than the best algorithm at the time. Then, the tradeoff disappeared in 1982 with Kadane's algorithm, which is still the single optimal algorithm for this problem.
For the rest of the 117 problem families' Pareto frontier graphs, they are available on our \href{https://algorithm-wiki.csail.mit.edu}{Algorithm Wiki website}.

In our analysis, we find that currently 17\% of the problem families in our database have time-space tradeoffs. The rest of the problem families have an algorithm that has a better (or equal) time and space complexity than the rest of the algorithms for that problem.

We also observed that over time, the share of problem families with such a tradeoff has been increasing by 1.79 percentage points per decade, as new algorithms are published. This historical trend is shown in Figure~\ref{subfig:tradeoff_history}. It is unclear what causes this increase, as it could be due to researchers developing new algorithmic tools that require non-trivial tradeoffs, or it could occur from researchers running into inherent resource tradeoffs needed to solve these problems.

\section{Conclusion}\label{sec:conclusion}
In past work, the time complexity of algorithms has been studied intensively to measure the speed of computational progress. In this work, we focus on space complexity, an area of growing importance that has been widely neglected in much of algorithm literature in the past. As part of this work, we have gathered the space complexities of over 800 algorithms, the vast majority of which we derived because the literature had not.

We show that, for ~20\% of algorithm problems, improvements to space complexity outpaced hardware improvements in DRAM access speed and DRAM capacity improvements. However, for the vast majority of problems (~80\%), space complexity has not seen any improvements because either the original algorithms were already space-efficient, or it is simply difficult for algorithm researchers to improve on the space complexity. It may also be that researchers simply care much less about improving space complexities than progressing other areas of algorithmic research, as suggested by the low rate of papers that analyze the space complexity of their algorithms. Thus, it would also be very interesting to know why that rate has been increasing over the years.

Although there have been impressive improvements in algorithmic space complexity, they are substantially smaller than the improvements in time complexity for the algorithms examined. This relative performance change means algorithmic progress may be a contributing factor to the memory wall phenomena. 

Having gathered this first comprehensive look at the space complexity of algorithms, we are making it easily accessible to the algorithm community via our website \href{https://algorithm-wiki.csail.mit.edu}{Algorithm Wiki}. We envision that the combination of this article's summarization and that website's comprehensive coverage will provide several benefits to the community. We hope this work will help guide theorists towards promising areas where they could come up with more space efficient algorithms. In addition it can help computer scientists implementing algorithms by supplying them with more information that they may use to help decide which algorithm is the most appropriate to use - particularly when there are time-space trade-offs among the best algorithms for the problem.

Overall, the broad scope and comprehensiveness of our study allows us to see for the first time the growing importance of space complexity and to quantify how it is evolving over time.

\section*{Acknowledgments}

We would like to acknowledge Khaleel al-Adhami who assisted in the gathering and derivation of space complexities for this paper.

We would also like to acknowledge Matrin Farach-Colton and Erik Demaine for useful discussion and feedback regarding different models of computation. In particular, Martin pointed out that space complexity can be larger than time complexity, and with Erik, we worked out the standard models in which that holds. Also, Michael Coulombe, one of Erik's students, helped us find a specific type of Turing machine, the log-space transducer, which we use in our discussion of constant auxiliary space in Word RAM.

We would also like to thank Ryan Williams and Virginia Vassilevska Williams for their insight into specific problems' space complexities, especially APSP and matrix multiplication.

We would like to thank Andrew Lucas for help with designing and maintaining the Algorithms Wiki website.

\bibliography{main}\label{sec:bibliography}
\bibliographystyle{plain}

\appendix

\section{Appendix}\label{sec:appendix}

\subsection{Additional Background}\label{sec:background}

In computer science, an \emph{algorithmic problem} is defined as a relation\footnote{In some cases problems will be defined as functions for mathematical convenience; however, many real computational problems have multiple acceptable answers.} from a set of values called \emph{inputs} to a set of values called \emph{outputs}, and an \emph{algorithm} is defined as a (finite) sequence of instructions that are used to solve an algorithmic problem. For example, integer addition can be thought of as an algorithmic problem, where the input consists of a set of two integers, and the output is another integer. In many cases, an input will correspond to exactly one or even multiple correct outputs. There are cases, however, where the output of a problem is not well defined. For example, with the problem `texture synthesis', which is the problem of trying to generate an image that looks like it has a given texture, there is no agreed-upon way to determine exactly whether the output image is correct or not. We consider problems like this to be ``inexact problems", and they are not included in our analysis. Algorithms can be deterministic or randomized, and in the case of randomized algorithms, we often consider the probability that inputs get mapped to a correct output. In this work, we exclude randomized algorithms from our analysis since it is not straightforward to compare them with deterministic algorithms in a way that also incorporates the probability that the randomized algorithm produces a correct output.

In the analysis of algorithms, two fundamental metrics of algorithmic efficiency are \emph{time complexity} and \emph{space complexity}. These consider how long it takes an algorithm to produce an output, and how much information an algorithm needs to store at a time while performing its computation. To make our algorithms more efficient, we want to reduce the amount of time and storage our algorithm uses while solving a problem.

We usually discuss these metrics with respect to the size of the input, and focus on when the size of the input is large or tends to infinity. As such, we use \emph{asymptotic notation} when discussing time and space complexities, ignoring any constant factors (i.e. factors that do not change when the size of the input changes) and lower-order terms. For example, if an algorithm takes at most $\frac{n^2-n}{2}$ steps and uses $2n$ units of space to solve a problem, we say that the time complexity is $O(n^2)$, since we ignore the constant factor of $\frac {1}{2}$ and the lower-order term $-n$, and the space complexity is $O(n)$, since we ignore the constant factor of $2$.

More details about how we measure time and space complexity, and what models of computation are used can be found in Section~\ref{sec:methods}.

\subsection{Problem Families with Super-linear Best Space Complexity Algorithms}\label{sec:superlinear_space}

The following problem families have no algorithms in our database with auxiliary space complexity that is less than or equal to linear. They are organized by their most space-efficient algorithm's auxiliary space complexity classification.
The specific definitions of these problems as well as all of the other problem families we studied can be found on our \href{https://algorithm-wiki.csail.mit.edu}{Algorithm Wiki website}.
\begin{itemize}
    \item Quadratic auxiliary space
    \begin{itemize}
        \item Maximum Flow
        \item All-Pairs Shortest Paths (APSP)
        \item Integer Relation
        \item Maximum-Weight Matching
        \item Constructing Eulerian Trails in a Graph
        \item Poisson Problem
        \item CFG Problems
        \item Finding Frequent Itemsets
        \item Graph Isomorphism Problem
        \item Determinant of Matrices with Integer Entries
        \item Transitive Reduction Problem
        \item Secret Sharing
    \end{itemize}
    \item Cubic auxiliary space
    \begin{itemize}
        \item Longest Path Problem
    \end{itemize}
    \item Polynomial ($>$ 3) auxiliary space
    \begin{itemize}
        \item Traveling-Salesman Problem (TSP)
    \end{itemize}
    \item Exponential auxiliary space
    \begin{itemize}
        \item Informed Search
        \item Gröbner Bases
        \item Coset Enumeration
        \item Dependency Inference Problem
        \item Frequent Words with Mismatches Problem
    \end{itemize}
\end{itemize}

\end{document}